\documentclass{article}
\usepackage[dvips]{graphicx}
\usepackage{amssymb}
\usepackage{float}
\usepackage{authblk}
\usepackage{url}
\usepackage{cite}

\title{\huge\textbf{Some optical and dynamical}
\huge\textbf{phenomena in the Rindler}
\huge\textbf{model}\linebreak
\small{refereed version of this article: Open Journal of Modern Physics, Vol. 1, No. 1, pp. 8-12, 2014}}

\author[1]{Emrah Birsin \thanks{Email: emrah@physik.hu-berlin.de}}
\author[2,3]{Wolfgang Hasse \thanks{Email: astrometrie@gmx.de}}
\affil[1]{Humboldt University of Berlin, Department of Physics, Newtonstr. 15, 12489 Berlin, Germany.\newline}
\affil[2]{Technical University Berlin, Institute for Theoretical Physics, Sekr. EW 7-1, Har\-denbergstr. 36, 10623 Berlin, Germany.\newline}
\affil[3]{Wilhelm Foerster Observatory Berlin, 
Munsterdamm 90, 12169 Berlin, Germany.}

\begin{document}
\maketitle

\begin{abstract}
In Rindler's model of a uniformly accelerated reference frame we
 analyze the apparent shape of rods and marked light rays for the case that the
 observers as well as the rods and the sources of light are at rest with respect to the Rindler
 observers. Contrary to the expectation suggested by the strong
 principle of equivalence, there is no apparent ''bending down'' of a light ray
 with direction transversal to the direction of acceleration, but a straight rod oriented orthogonal to the direction of
 acceleration appears bended ''upwards''. These optical phenomena are in accordance with the dynamical experience of
 observers guided by a straight track or a track curved in the same way as the marked
 light ray, respectively: While the former observer feels a centrifugal force
 directed "downwards", the centrifugal force for the latter vanishes. The properties of gyroscope transport along such tracks are correspondingly. \newline\newline
\end{abstract}

\section{INTRODUCTION}
Rindler's \cite{Rindler1977} well known model consists of a set of observers who are accelerating at
 constant rate and direction in Minkowski spacetime (i. e. the acceleration vector is Fermi-propagated
 along the observer's worldline), such that they are mutually at rest (tangent vectors of the worldlines are parallel along the rest spaces).
 Consequently the rate of acceleration is not the same for all observers.

 Among the further publications on uniformly accelerated reference fra\-mes there are some papers discussing several aspects of
 light propagation and optical effects. So Hamilton \cite{Hamilton1978} analyzed the apparent diameter of an object in free fall
 as well as the gravitational redshift and the occurrence of an ''event horizon'', which has already been pointed out by
 Rindler \cite{Rindler1966}. The latter two phenomena had also been treated by Good \cite{Good1982} and Desloge and Philpott
 \cite{EA1987}. Owing to the rapid progress in computer technology in the last two decades, sophisticated numerical visualization methods for several special relativistic effects have been developed (for a recent survey see \cite{Weiskopf2010}). But the obvious question regarding the effect of the acceleration on the optical appearance of straight rods
 and marked light rays seems to be open (at least regarding a solution by analytical methods), although it should be relevant not least in view of the strong principle of equivalence. And what is more, in teaching relativity it is mostly, without any doubt, assumed that the light bending in a gravitational field or an accelerated elevator is, in principle, visible to an observer, see, e. g., the lectures \cite{Daw2013}.

 In the present paper we aim to fill this gap. To this end we introduce in the following section
 our concepts of the two kinds of objects before we calculate their apparent shape in the third section. The fourth section
 analyzes the relation of the optical effects to inertial forces. The discussion of our results in the last section contains
 among others an explanation of the optical effects in terms of retardation effects.

\section{THE MODEL AND COORDINATES}
Let us consider standard coordinates $x, y, z, t$ in Minkowski
spacetime restricted to the ''Rindler wedge'' ($z>|ct|$) which is equipped with a family of worldlines of
 constant proper acceleration $a$ in $z$-direction given by the hyperbolas (Rindler observers)

$$
\label{hyperbola} z^2-c^2t^2=\frac{c^4}{a^2}
$$
($x$ and $y$ are constant).

Our objects of observation are \newline
\newline
(A) \parbox[t]{13cm}{a straight rod (infinitely
 long) in $x$-direction at rest relative to the \\ Rindler observers and}
\newline \newline \newline
(B) \parbox[t]{13cm}{a marked light ray (infinitely long), emitted
orthogonal to the $y$-\\ direction by a source which is at rest relative to the
Rindler\\  observers (''marked'' means, that the light is partially
scattered,\\ e.g. by steam or dust particles, so that the light ray is visible to\\ all observers).}
\newline
\newline

The restriction to rods oriented in $x$-direction avoids some conceptual problems concerning the notion of a straight accelerated rod:
A rod (modelled by a two-dimensional timelike surface) which is straight in any inertial system (in the sense that the intersections
 of the surface with the rest spaces are straight lines) cannot be transversally accelerated (i. e., the surface is a plane).
Our rods can be simply constructed by a rod initially at rest in an inertial system and an instantaneous (w. r. t. this system) start
 of the acceleration of all its parts.

The two kinds of objects are modelled by two-dimensional surfaces
(''worldsheets''). In both cases we construct the worldsheet of
the object by dragging of a suited straight line with the flow of
the boost Killing vector field. These straight lines are given by
\newline
 \newline
(A) \hspace{1cm} \parbox[t]{12cm}{$ x=\lambda, \qquad y=y_r, \qquad z=z_r,
\qquad t=0$
\newline (the rod at $t=0$, $\lambda$: parameter along the rod) and}
\newline \newline
(B)\hspace{1cm} \parbox[t]{12cm}{$ x=\lambda\cos\alpha,\qquad y=y_s,\qquad z=z_s+\lambda\sin\alpha,
\qquad ct=\lambda
$ \newline(the worldline of the light signal emitted at $t=0$, position of the\\ source at $t=0$:
$x=0$, \ $y=y_s$, \ $z=z_s$; \ $\lambda$ : parameter along the\\
light signal, the angle $\alpha$ gives its direction in the $x$-$z$-plane).}
\newline
The dragging mentioned above is
given by the equations\newline
\newline(A)
\begin{equation}
\label{rodtapematrix} \left(\matrix{x \cr y\cr z \cr ct}\right) = \left(\matrix{1&0&0&0 \cr
 0&1&0&0 \cr 0&0&\cosh{\eta}&-\sinh{\eta}\cr 0&0&-\sinh{\eta}
 &\cosh{\eta}}\right)\left(\matrix{\lambda \cr y_r \cr z_r \cr
 0}\right) = \left(\matrix{\lambda \cr y_r \cr z_r\cosh{\eta} \cr
 -z_r \sinh{\eta}}\right)
\end{equation} \newline
(B)
$$
\left(\matrix{x \cr y\cr z \cr ct}\right) = \left(\matrix{1&0&0&0 \cr
 0&1&0&0 \cr 0&0&\cosh{\eta}&-\sinh{\eta}\cr 0&0&-\sinh{\eta}
 &\cosh{\eta}}\right)\left(\matrix{\lambda \cos{\alpha} \cr y_s \cr z_s + \lambda \sin{\alpha}\cr
 \lambda}\right) =
$$ \newline
\begin{equation} \label{raytapematrix}
= \left(\matrix{\lambda \cos{\alpha} \cr y_s \cr (z_s+\lambda\sin{\alpha})\cosh{\eta}-\lambda\sinh{\eta} \cr
-(z_s+\lambda\sin{\alpha}) \sinh{\eta}+\lambda\cosh{\eta}}\right),
\end{equation} \newline \newline
where $\eta$ is the boost parameter.

\begin{figure}[H]
\includegraphics[scale=0.45]{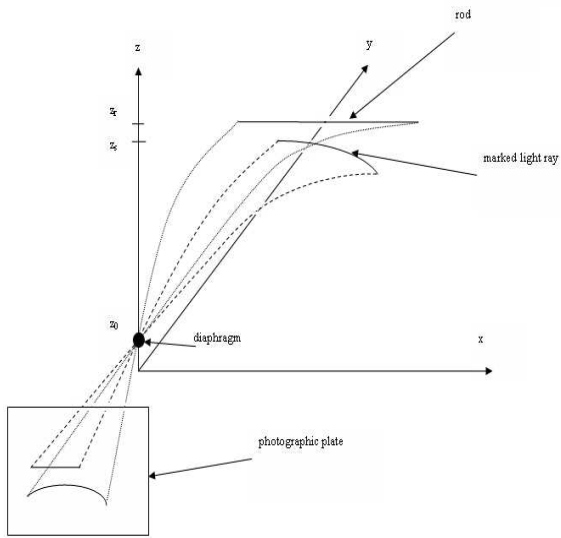}
\caption{Spatial arrangement of the objects and the observer}
\end{figure}
Our observer is a Rindler observer, located at the event $x=0$, $y=0$,
$z=z_0$, $t=0$, whose past light cone is given by
\begin{equation}
\label{cone} x^2+y^2+(z-z_0)^2-c^2 t^2=0.
\end{equation}

The set of parameters ($y_r$, $z_r$, y$_s$, $z_s$, $\alpha$, $ z_0$) covers all possible spatial arrangements of the objects
 relative to the observer (see Figure 1, concerning the $x-$separation of light ray and observer, we have to make use of suitable
 combinations of $z_s$ and $\alpha$). Note that due to the invariance under the flow of the boost Killing field it is
 sufficient to consider one moment of observation, say at $t=0$, and
 that the instantaneous acceleration of the observer is irrelevant for the apparent shape
 of the objects (which is determined by the spatial directions of the incoming light signals; if the "light bending" within
 the observer's optical imaging system is not negligible, one has to consider an inertial observer instantaneously at rest
 w. r. t. our Rindler observer), but a velocity relative to the Rindler observers would give
 rise to aberration, and with that to an apparent distortion of the objects. \newline

\section{APPARENT SHAPE OF RODS AND \\ LIGHT RAYS}

In order to calculate the apparent shapes we have to consider the events at the worldsheets of the rod and the light ray, respectively, which are connected
 with the event of observation by light signals. To that end let us eliminate the parameters $\lambda$ and $\eta$ from (\ref{rodtapematrix}) and
(\ref{raytapematrix}), resulting in \newline
(A)
\begin{equation}
\label{rodtape} y=y_r, \qquad z^2-c^2 t^2=z_r^2
\end{equation}
(hyperbolic bended tape in spacetime) and \newline
(B)
\begin{equation}
\label{raytape} y=y_s,  \qquad x^2+z^2-c^2 t^2=z_s^2+2z_sx\tan\alpha
\end{equation}
(one-shell hyperboloid in spacetime centered at $x=z_s\tan\alpha$, $y=y_s$, $z=0$, $t=0$). In the next step we have to consider the intersection
of the cone (\ref{cone}) with the surfaces (\ref{rodtape}) and (\ref{raytape}),
 respectively, which leads to the equations:\newline (A)
\begin{equation}
\label{conerodintersection} x^2+y_r^2+z_r^2-2zz_0+z_0^2=0, \qquad
y=y_r, \qquad z^2-c^2t^2=z_r^2
\end{equation} (B)
\begin{equation}
\label{conerayintersection} 2xz_s\tan\alpha+y_s^2+z_s^2-2zz_0+z_0^2=0, \qquad
y=y_s, \qquad x^2+z^2-c^2t^2=z_s^2
\end{equation}

The optical observation can be realized by a central projection through a lens inserted in a small diaphragm
 onto the retina or a photographic plate in the focal plane of the lens. In the case of a plane plate
 perpendicular to the $y$-direction the image of the
 object on the plate is in the limit of an infinitesimally small optical imaging system (cf. the remark in parantheses in the last paragraph of Section 2), up to a scaling factor and a rotation about the angle
 of $\pi$, exactly (!) given by "forgetting" the equations containing $t$ and $y$
 in (\ref{conerodintersection}) and (\ref{conerayintersection}) (projection in the
 observer's infinitesimal rest space followed by a projection onto the $x$-$z$-plane).
 The remaining equations describe a bended upwards parabolic curve and a straight line,
 respectively, in the $x$-$z$-plane which may be identified with the photographic plate
 (or with the retina, as far as its curvature is negligible).

So we have shown that the rod appears sagging while the marked light ray looks unbended. (The marked light ray corresponds to a geodesic on the
observers celestial sphere.)

\section{INERTIAL FORCES}
One may argue that our result of the apparent bending of a straight rod is
 contrary to the dynamical experience of an observer who is moving along the
 rod. If the optical impression does not delude, he or she should feel a
 centrifugal force antiparallel to the direction of acceleration. If we imagine
 an elevator whose acceleration is constant in the course of time, our results say that its bottom
 appears boiled downwards, so a ball rolling along the bottom should exert a
 force on it which is greater than in the case of a ball lying at rest on the
 bottom (even in the respective rest frame of the ball, so we must not appeal to
 the relativistic increase of mass with velocity). This does not fit in our
 image of the Rindler model as a realization of an as far as possible homogeneous
 field of gravitational force.
 
But indeed, this centrifugal force effect exists in the Rindler model, as we can
 show with reference to the results of the work of Foertsch, Hasse and
 Perlick \cite{HassePerlick} on inertial forces in arbitrary spacetimes.
 
 Newton's concept of centrifugal force attributes it to the rotation of a reference frame (with, at
 least in classical theory, respect to the global rest frame) whereas Huygens' definition attributes
 it (in the "corotating" reference frame) to the motion along a curved trajectory. While in
 Newtonian theory these two concepts are equivalent, they are not compatible in general
 relativity as has been pointed out by Abramowicz \cite{Abramo}. He had demonstrated
 that, under the two alternatives, only Huygens' definition is consistent with general relativity.
 
Foertsch et al. \cite{HassePerlick} share Abramowicz's view and
give definitions of inertial forces as a literal adaption of
Huygens' definition to general relativity. They consider in an
arbitrary spacetime $(M,g)$ a two-dimensional timelike submanifold
$\Sigma$, which models a track in the course of time, and an
observer field $n$ on $\Sigma$. A particle moving along the track is given by a timelike curve on $\Sigma$ with tangent vector field $u$.
Furthermore they define the vector
field $\tau$ on the track (uniquely up to sign) by the conditions
 $g(n,\tau)=0 \;$ and $\; g(\tau ,\tau )=1$ ($n$ and $u$ are normalized by
 $g(n,n)=g(u,u)=-1$). Gravitational, centrifugal, Coriolis
and Euler forces are assigned to every particle worldline
$\lambda$ in $\Sigma$ with respect to $n$ by a decomposition of
the the part of the inertial acceleration $a=-\bigtriangledown_uu$
which is orthogonal to $n$ into parts orthogonal and parallel to
$\tau$, sorted by the powers of the velocity $v$ of the particle
with respect to $n$. Their central result consists in a theorem
which states that centrifugal and Coriolis forces vanish, for all
$\lambda$ in $\Sigma$ with respect to any $n$, if and only if
$\Sigma$ is a photon 2-surface, i. e., generated by two families of
lightlike geodesics.

Coming back to our special problem in Rindler's model, let us be given a straight track oriented in the
$x$-direction and at rest with
 respect to the Rindler observers. Now $n$ is given by the velocity field of the
 Rindler observers on the track in spacetime (bended tape, given by
 equations (\ref{rodtape})), the vector field $\tau$ is simply $\tau=\pm\frac{\partial }{\partial x}$
 and $u$ is the velocity vector of a particle guided by the track.
Since obviously $\bigtriangledown_n\tau=0$, $\bigtriangledown_\tau
n=0$, $\bigtriangledown_\tau \tau=0$ and $g(\bigtriangledown_n
n,\tau)=0$ the  equations (6) and (7) in the paper of Foertsch et
al. \cite{HassePerlick} yield a vanishing Coriolis force and a
centrifugal acceleration given by $a_{cen}=-\gamma^2 \beta^2
\bigtriangledown_n n$, where $\gamma$ denotes  the Lorentz
factor $\frac{1}{\sqrt{1-\beta^2}}$ ($\beta$: velocity of
the particle relative to the track in units of the velocity of
light). This centrifugal acceleration completes the
"gravitational" acceleration $a_{gra}=-\bigtriangledown_n n$ to
the total inertial acceleration
$$ a_{gra}+a_{cen}=-\gamma^2
\bigtriangledown_n n
$$
(apart from a possible Euler acceleration which is present if $\beta$ changes in
 the course of time, cf. (\ref{conerayintersection}) in \cite{HassePerlick}).

 On the other hand,
 if a track is curved in the same way as the marked light ray, not only the
 Coriolis force, but also the centrifugal force vanishes, in accordance with the
 optical image. This becomes clear if we consider light rays which are directed
 "backwards" the track additional to the "forward" ones. So
 we recognize that the two-dimensional surface of the track in spacetime is
 generated by two families of lightlike geodesics, i. e., it is a photon surface in
 the sense of Foertsch et al. \cite{HassePerlick}. In fact, since the worldsheet of the marked ray is part of the
 hyperboloid (\ref{raytape}), it is exactly Example 5.1 of a photon surface in \cite{HassePerlick}.

\section{DISCUSSION}

In Rindler's model of an uniformly accelerated reference frame we
 analyzed the apparent shape of rods and marked light rays for the case that the
 rods and the sources of light are at rest with respect to the Rindler
 observers.

It turned out that, contrary to the expectation suggested by the strong
 principle of equivalence, there is no apparent "bending down" of a light ray
 with direction transversal to the direction of acceleration.

Since, as another
 of our results, a straight rod oriented orthogonal to the direction of
 acceleration appears bended ''upwards'', there is nevertheless an apparent
 "bending down" of the ray relative to straight rods.

One may ask for the reason of the apparent "sag" relative to the expected shape for both kinds of objects. We can interpret
 it as a retardation effect: The parts of the objects which are off the
 optical axis are seen in an earlier position as the parts on the axis due to longer light travel time. In the instantaneous inertial
 system of the observer the  elevator (and the objects with it) moves downwards in the additional travel time of light, so the "outer"
 parts of the objects are in an apparent higher position.
 In the case of the marked light ray this retardation effect compensates exactly its real bending.

Alternatively, both optical phenomena are also understandable in terms of the Fermat geometry (optical reference geometry) of the Rindler model.
As we know \cite{Perlick}, its Fermat geometry is exactly the geometry of Poincare's half-space, whose geodesics are half circles with center at
 $z = 0$ as it is considered in figure 1 (Poincare's half-space is in turn globally isometric to Lobachevsky space).

Poincare's half-space gives an example for a space of constant curvature. As a matter of fact, just the spaces of constant curvature are characterized by the apparent straightness of marked light rays. More precisely formulated: In the conformally stationary case appears every marked light ray as a straight line for every observer at rest with respect to the reference frame given by the conformal Killing vector field if and only if the Fermat geometry describes a space of constant sectional curvature with vanishing Fermat two form. Because this result goes somewhat beyond our questions concerning the Rindler model, we have separated the proof out in an appendix and leave it by a rough draft.

Furthermore we proved that these optical phenomena are in accordance with the dynamical experience of
 observers guided by a straight track or a track curved in the same way as the marked
 light ray, respectively: While the former observer feels a centrifugal force
 directed "downwards", i. e. antiparallel to the acceleration of the reference
 frame, the centrifugal force for the latter vanishes. Whereas in the classical theory the worldsheet of a rod accelerated in z-direction
 is invariant under the Galilei boosts in x-direction, in special relativity there is no invariance under the respective Lorentz boosts.
 Therefore one can speak of an absolute motion along the rod which manifests itself in a centrifugal force.

 The theorem of Foertsch et al. \cite{HassePerlick} ensures also,
that the optical phenomena in the Rindler
 model are in accordance with the properties of gyroscope transport along the tracks.
The apparent "sag" of the rod corresponds to a rotation of the axis of a gyroscope guided along the track.
  If the axis is initially parallel
 to the track, it will rotate downwards relative to the track. But in the case of the marked light ray there is no such rotation.

Because of the strong principle of equivalence, there must be similar effects in the gravitational field of a massive body,
e. g. on the surface of a neutron star, whose outer region might be described by
Schwarzschild spacetime. Usually the reference to the strong principle of equivalence rests on the imagination of a homogeneous
gravitational field. But in view of the apparent bending of straight rods and the centrifugal force,
apart from the variation of the acceleration with the height, the Rindler model meets hardly the requirements on homogenity.
Furthermore, in order to transfer our results to ''genuine'' gravitational fields, we need an ''identification'' of some part of Schwarzschild spacetime with part of the Rindler model (on certain similarities of the Rindler model to Kruskal space see \cite{Rindler1966}), but such an ''identification'' cannot result from the limit of great distances from the event horizon, as can be seen by the different directions of the centrifugal forces. (Moreau et al. \cite{Moreau} give a splitting of the Schwarzschild line element into a part corresponding to an accelerated reference frame in flat spacetime and a part related to curvature, but this splitting is restricted to a small spacetime region around some origin and, moreover, turns out to be coordinate-dependent.) These problems will be the subject of a forthcoming paper.

\section*{Appendix}
\subsection*{Fermat geometries with apparently straight light rays}
In the framework of the Fermat geometry the light rays are given by the solutions $\hat{\gamma}$ of the second-order differential equation
\begin{equation}
\label{sode} \hat{g}(\hat{\triangledown}_{\dot{\hat{\gamma}}}\dot{\hat{\gamma}},\cdotp)=2\hat{\omega}(\cdotp,\dot{\hat{\gamma}})
\end{equation}
where $\hat{g}$ and $\hat{\omega}$ denote the Fermat metric and the Fermat form, respectively, defined on $\hat{M}$, the set of all integral curves of the conformal Killing field (''the three-dimensional space''), and $\hat{\triangledown}$ is the Levi-Civita connection belonging to $\hat{g}$, see Perlick \cite{Perlick}. The property of a light ray to appear as a straight line to some observer located in a point $\hat{p}_0\in\hat{M}$ is characterized by the condition that the one-parameter family of light rays connecting the given ray with $\hat{p}_0$ has tangent vectors which intersect the unit sphere of the tangent space $T_{\hat{p}_0}\hat{M}$ in a common great circle. Thus we have to prove the following.

\textbf{Theorem} The condition described above holds for every pair consisting of a lightray and a point in $\hat{M}$ if and only if $(\hat{M},\hat{g})$ is a space of constant sectional curvature and $\hat{\omega}=0$.

\textbf{Outline of the proof}\newline
Since the ''if'' part is plausible by symmetry arguments (in a space of constant curvature there is no preferred spatial direction, therefore the optical image of a marked ray cannot deviate from a straight line by bending in any direction) we confine ourself to the ''only if'' part. Furthermore, if the condition of the apparent straightness holds true it is strongly suggested that $\hat{\omega}=0$: Note that in the case of a non-vanishing $\hat{\omega}$ the shape of a light ray connecting two given points in $\hat{M}$ changes by interchanging the light source and its target (cf. eq. (\ref{sode})). At least for some subset of the observers the two components of such a pair of rays must be separated on the celestial sphere, in contradiction to the condition that all images of light rays are great circles at the celestial sphere of any observer. Therefore, we here omit the technical details of proving $\hat{\omega}=0$.

It remains to prove that under the condition of the apparent straightness, which is by virtue of $\hat{\omega}=0$ a property of the $\hat{\triangledown}$ geodesics alone, the space $(\hat{M},\hat{g})$ is locally maximally symmetric. To that end we consider two-dimensional submanifolds generated by a geodesic (the ''marked ray'') and all geodesics connecting it with a given point (''view lines'' of the observer). By arguments based on the uniqueness of the solutions of the geodesic equation for given initial values it is an easy task to show that any geodesic (and not only the special ones used for the construction) with starting point in the considered submanifold and starting vector tangent to it extends entirely in it; in other words, any such submanifold is totally geodesic. Now let us be given an arbitrary point $\hat{p}\in\hat{M}$ and four tangent vectors $X,Y,Z,W\in T_{\hat{p}}\hat{M}$ with $X,Y$ and $Z$ complanar, defining a two-dimensional subspace $S_{\hat{p}}$, and $W\perp S_{\hat{p}}$. Since by means of the exponential map any such subspace can be obtained as the tangent space of a suitable submanifold constructed as above and containing $\hat{p}$, we can convert the condition of the vanishing of the second fundamental form of the submanifold in $\hat{p}$ (which characterizes the property of being totally geodesic) to the condition
\begin{equation}
\label{condi} \hat{g}(\hat{R}(X,Y)Z,W)=0
\end{equation}
($\hat{R}$ denotes the Riemannian curvature tensor of the connection $\hat{\triangledown}$) for any quadruple of vectors of the given form (eq. (\ref{condi}) results by introducing $\hat{R}$ as the commutator of the second covariant derivatives, the latter turn out to be orthogonal to W by vanishing of the second fundamental form).

Finally, with some elementary linear algebra we can deduce from (\ref{condi}) that the Ricci tensor $\hat{Ric}$ is related to the curvature scalar $\hat{S}$ by $\hat{Ric}=\frac{\hat{S}}{3}\hat{g}$ ($\hat{S}$ be found constant on $\hat{M}$ by applying the contracted Bianchi identity). In the three-dimensional case this is sufficient for maximal local symmetry (space of constant sectional curvature $\hat{K}=\frac{\hat{S}}{6}$). \qquad \qquad \qquad\qquad \qquad \quad $\square$

The Fermat geometry of the Rindler model is given by the special case $\hat{K}=-1$. The fact that a maximally symmetric geometry is related to the Rindler model stands in contradiction to the suggestion evoked by the preferred ''downwards'' spatial direction, which is however, as we have seen, really not be ''felt'' by the light alone (but, of course, by combined light/ordinary-matter systems as, e. g., a light ray in an accelerated elevator).

\bibliography{./rindlerBib}
\bibliographystyle{hunsrt}
\end{document}